\documentclass[]{aastex631}

\usepackage[whole]{bxcjkjatype}

\newcommand{\twelvecoh}{\mbox{$^{12}$CO($J$ = 2--1)}} 
 
\newcommand{\thirteencoh}{\mbox{$^{13}$CO($J$ = 2--1)}} 
\newcommand{\twelvecol}{\mbox{$^{12}$CO($J$ = 1--0)}}

\begin{document}

\title{A multi-scale molecular and atomic gas view on the H$\;${\sc ii} region N113 in the Large Magellanic Cloud: Evidence for high-mass star formation triggered by supersonically-colliding H$\;${\sc i} flows}

\author[0000-0002-1865-4729]{Rin I. Yamada}
\affiliation{National Astronomical Observatory of Japan, National Institutes of Natural Sciences, 2-21-1 Osawa, Mitaka, Tokyo 181-8588, Japan}
\affiliation{Faculty of Engineering, Gifu University, 1-1 Yanagido, Gifu 501-1193, Japan}
\affiliation{Department of Physics, Nagoya University, Furo-cho, Chikusa-ku, Nagoya 464-8601, Japan}

\author[0000-0002-2062-1600]{Kazuki Tokuda}
\affiliation{Faculty of Education, Kagawa University, Saiwai-cho 1-1, Takamatsu, Kagawa 760-8522, Japan}
\affiliation{Department of Earth and Planetary Sciences, Faculty of Science, Kyushu University, Nishi-ku, Fukuoka 819-0395, Japan}
\affiliation{National Astronomical Observatory of Japan, National Institutes of Natural Sciences, 2-21-1 Osawa, Mitaka, Tokyo 181-8588, Japan}

\author[0000-0002-8966-9856]{Yasuo Fukui}
\affiliation{Department of Physics, Nagoya University, Furo-cho, Chikusa-ku, Nagoya 464-8601, Japan}

\author[0009-0000-0359-9563]{Daiki Adachi}
\affiliation{Department of Physics, Graduate School of Science, Osaka Metropolitan University, 1-1 Gakuen-cho, Naka-ku, Sakai, Osaka 599-8531, Japan}

\author[0000-0003-2248-6032]{Marta Sewi{\l}o}
\affiliation{Exoplanets and Stellar Astrophysics Laboratory, NASA Goddard Space Flight Center, Greenbelt, MD 20771, USA}
\affiliation{Department of Astronomy, University of Maryland, College Park, MD 20742, USA}
\affiliation{Center for Research and Exploration in Space Science and Technology, NASA Goddard Space Flight Center, Greenbelt, MD 20771, USA}

\author[0000-0002-4663-6827]{R\'{e}my Indebetouw}
\affiliation{Department of Astronomy, University of Virginia, PO Box 400325, Charlottesville, VA 22904, USA}
\affiliation{National Radio Astronomy Observatory, 520 Edgemont Road, Charlottesville, VA 22903, USA}

\author[0000-0002-3925-9365]{C.-H. Rosie Chen}
\affiliation{Max Planck Institute for Radio Astronomy, Auf dem Huegel 69, D-53121 Bonn, Germany}

\author[0000-0002-2794-4840]{Kisetsu Tsuge}
\affiliation{Institute for Advanced Study, Gifu University, 1-1 Yanagido, Gifu 501-1193, Japan}
\affiliation{Faculty of Engineering, Gifu University, 1-1 Yanagido, Gifu 501-1193, Japan}

\author{Takeru Nishioka}
\affiliation{Department of Physics, Nagoya University, Furo-cho, Chikusa-ku, Nagoya 464-8601, Japan}

\author[0000-0003-2062-5692]{Hidetoshi Sano}
\affiliation{Faculty of Engineering, Gifu University, 1-1 Yanagido, Gifu 501-1193, Japan}
\affiliation{Center for Space Research and Utilization Promotion (c-SRUP), Gifu University, 1-1 Yanagido, Gifu 501-1193, Japan}

\author{Mao Tamashiro}
\affiliation{Department of Physics, Nagoya University, Furo-cho, Chikusa-ku, Nagoya 464-8601, Japan}

\author[0000-0001-8901-7287]{Naslim Neelamkodan}
\affiliation{Department of Physics, United Arab Emirates University, Al-Ain, 15551, UAE}

\author[0000-0002-7759-0585]{Tony Wong}
\affiliation{Astronomy Department, University of Illinois, Urbana, IL 61801, USA}

\author{Lynn R. Carlson}
\affiliation{Department of Science, College of Southern Maryland, 8730 Mitchell Rd, PO Box 910, La Plata, MD 20646}

\author[0000-0002-0861-7094]{Joana M. Oliveira}
\affiliation{Lennard Jones Laboratories, School of Chemical \& Physical Sciences, Keele University, Staffordshire ST5 5BG, UK}

\author[0000-0001-7813-0380]{Akiko Kawamura}
\affiliation{National Astronomical Observatory of Japan, National Institutes of Natural Science, 2-21-1 Osawa, Mitaka, Tokyo 181-8588, Japan}

\author[0000-0002-1411-5410]{Kengo Tachihara}
\affiliation{Department of Physics, Nagoya University, Furo-cho, Chikusa-ku, Nagoya 464-8601, Japan}

\author[0000-0001-7826-3837]{Toshikazu Onishi}
\affiliation{Department of Physics, Graduate School of Science, Osaka Metropolitan University, 1-1 Gakuen-cho, Naka-ku, Sakai, Osaka 599-8531, Japan}

\begin{abstract}
The Large Magellanic Cloud (LMC) exhibits vigorous high-mass star formation, including the H$\;${\sc ii} regions 30~Dor that is the most active site of star formation in the local group. The present paper focuses on the Giant Molecular Cloud (GMC) in the H$\;${\sc ii} region N113 in the central part of the LMC. Based on the $\twelvecoh$ and $\thirteencoh$ data at a resolution of approximately 0.2\,pc taken with ALMA+APEX, we reveal that the GMC consists of two filamentary structures each of approximately 10 pc in length, forming a V-shape pattern with a vertex angle of 90 degrees. The filamentary structures host high-mass young stellar objects in gravitationally bound dense gas. Large-scale H$\;${\sc i} gas data covering 100 pc reveal two distinct velocity components separated by more than 40\,km\,s$^{-1}$, that correspond to the low velocity (L-) and disk (D-) H$\;${\sc i} components of the LMC. The L-component appears to be located in a cavity-like distribution of the D-component, and the CO filaments are positioned at the cavity's edge. We find evidence for the L-component to fit the cavity by a 53\,pc displacement, and suggest that collisional compression of the H$\;${\sc i} gas during the last 1.3\,Myr triggered the GMC formation and the high-mass star formation. This lends support for the large scale collision driven by the tidal interaction is playing a role in evolution of interstellar medium in N113.

\end{abstract}


\section{Introduction} \label{sec:intro}
High-mass stars are influential in galaxy evolution both in the physical and chemical contexts by the action of ultraviolet radiation, strong stellar winds, supernovae, and injection of heavy elements. It is therefore a key issue in astronomy to understand the formation mechanism of high-mass stars. The Large Magellanic Cloud (LMC) shows active high-mass star formation, as demonstrated by the large H$\;${\sc ii} region 30 Dor and the other regions of high-mass star formation. The LMC is located at 50\,kpc \citep{Pietrzynski_2019} from the sun and is the nearest star-forming galaxy, allowing us to achieve high (sub-pc) spatial resolution.  In addition, it is nearly face-on \citep{Balbinot_2015}, and observations can be made without heavy contamination along a line of sight, allowing for a reliable comparison among multi-phase gas and stellar components traced in multi-wavelength measurements.

Recently, collisions between two molecular/atomic clouds (cloud--cloud collisions = CCC) have been recognized as an important process of high-mass star/cluster formation, and more than 50 regions are listed as CCC candidates (see \citealt{Fukui_2021} for a review). The CCC candidates in the Milky Way have collision velocities of the order of 10\,km\,s$^{-1}$, while the CCCs driven by the galactic interactions have larger collision velocities in a range of 20--150\,km\,s$^{-1}$. There is a general trend that the large collision velocity and large gas column density lead to the formation of higher-mass stars and clusters; the Galactic CCCs usually form 1--10 O-type stars (hereafter O star) with a maximum cluster mass of 10$^4$\,$M_{\odot}$ as found in W3, M42, NGC~3603, RCW~38, and Westerlund~2 \citep{Yamada2024, Fukui_2018,Fukui_2014,Fukui_2016,Furukawa_2009}, and the galactic interactions form clusters consist of more that 100 O star with cluster masses of 10$^5$--10$^6$\,$M_{\odot}$ as found in 30 Dor and the Antennae Galaxies \citep{Fukui_2017,Tsuge_2019,Tsuge_2021a,Tsuge_2021b}. The theoretical studies of CCCs are also being promoted extensively \citep[e.g.,][]{habe_1992,Inoue_2013,Takahira_2014,Takahira_2018,Haworth_2015,Inoue_2018,Sakre_2021,Sakre_2023,Maeda_2021,Abe_2022}. 

The observational and theoretical studies indicate that the LMC made a close encounter with the Small Magellanic Cloud (SMC) 0.2 Gyr ago, and the tidal interaction between the two galaxies stripped gas from each of them \citep{Fujimoto_1990, Bekki_2007}. The stripped H$\;${\sc i} gas afterwards falls to and collides with both galaxies in the order of 100\,km\,s$^{-1}$ to trigger extensive star formation. The three sites of high mass star formation, 30 Dor, N44, and N159, are suggested to be such sites of triggered star formation by CCCs in the LMC within a few Myr \citep{Fukui_2015,Fukui_2017,Fukui_2019,Saigo_2017,Tokuda_2019,Tokuda_2022N159,Tsuge_2019}. In particular, it deserves special mention that the magneto-hydrodynamical numerical simulations by \cite{Inoue_2018} provided valuable insight into the CCC process in the LMC; the model adopts the initial condition that a spherical cloud causes a head-on collision with an extended gas disk and compresses the gas and magnetic field into a cone-like distribution. 
The model explains well the distribution of the fan-shaped clouds with filamentary features as well as the high-mass star formation near the cone vertex in the three clouds N159E, N159W-S, and W-N \citep{Fukui_2019,Tokuda_2019,Tokuda_2022N159}. These clouds are located in the H$\;${\sc i} Ridge, which is H$\;${\sc i}-dense region extending $\sim$1 kpc (R.A.) $\times$ $\sim$3~kpc (Dec.) in the south-east part of the LMC\citep{Luks_1992, Fukui_2017}. It is therefore of interest to pursue how extensively the high-mass star formation is triggered by the colliding H$\;${\sc i} flows over the LMC. It is also important to understand such triggered star formation over the whole LMC to better elucidate the role of tidal triggers in the galaxy's evolution. Some recent papers are focusing on the overall distribution of young stellar objects (YSOs), O stars, the extinction, and the H$\;${\sc i} gas for the whole LMC \citep{Furuta_2019,Furuta_2021,Furuta_2022,Kokusho_2023,Tsuge_2024} aiming to reveal the impact of the tidal trigger on the interstellar medium (ISM) and star formation. 

\begin{figure}[htb!]
    \centering
    \includegraphics[width=1\columnwidth]{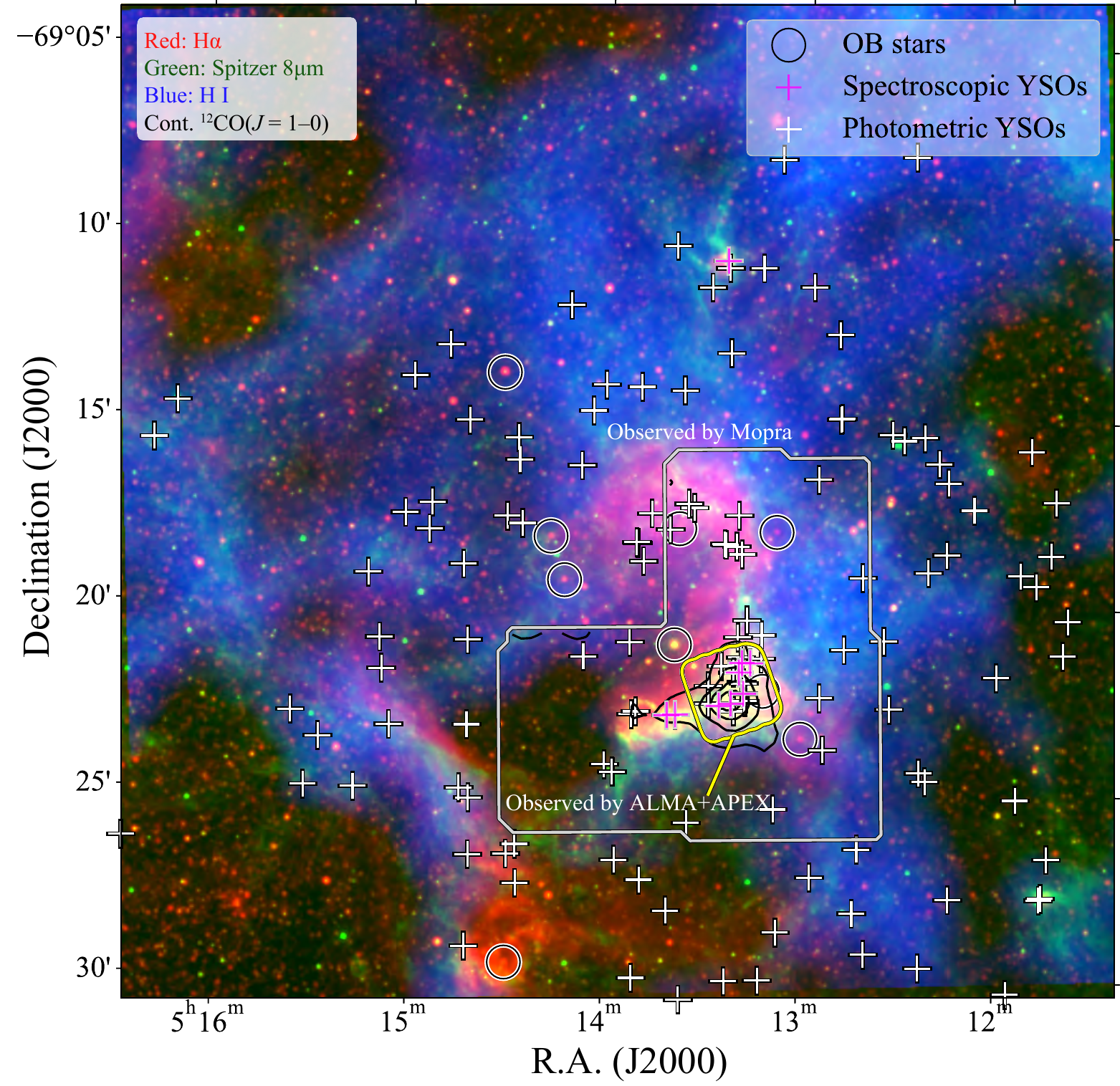}
    \caption{A large-scale view of the multi-phase interstellar medium toward the star-forming region N113. The three-color composite shows the MCELS H$\alpha$ (red; \citealt{Smith_1999}), Spitzer/IRAC~8\,$\mu$m (green; \citealt{Meixner_2006}), and H$\;${\sc i} (blue; \citealt{Kim_2003}) images. Black contours represent the CO (1--0) intensity integrated over the velocity range (-12.0, -23.0)~km~s$^{-1}$ obtained by Mopra \citep{Wong_2011, Wong2017}. Magenta crosses, white crosses, and black circles indicate the positions of spectroscopically-confirmed YSOs \citep{Seale_2009}, photometric YSOs \citep{Whitney2008, Gruendl2009, Carlson2012}, and WR/OB stars \citep{Bonanos_2009}, respectively. The green box shows the coverage of our ALMA+APEX mosaics.}
    \label{fig:1}
\end{figure}

In the present paper, we focus on N113, a massive star-forming region in the central Large Magellanic Cloud. We analyze the $J$~=~2--1 transition of carbon monoxide (CO) data taken with Atacama submillimeter/millimeter array (ALMA) combined with Atacama Pathfinder EXperiment (APEX) at a resolution of 0.2 pc, together with H$\;${\sc i} data on $\sim$100 pc scales surrounding N113 in order to reveal the high-mass star formation mechanism in N113. The overview of the region is shown in Figure~\ref{fig:1}, a three-color composite image of H$\alpha$ (ionized gas), 8\,$\mu$m PAH emission\footnote{PAHs are a type of molecule. Although it is still debated whether PAH emission accurately reflects the amount of H$_2$ molecules especially at smaller scale, it is certain that molecular gas exists in the direction of PAH emission \citep[see][and references therein]{Leroy2023}. Therefore, PAH emission is used to indicate the approximate distribution of low-density molecular gas that cannot be traced by CO.}, and H$\;${\sc i}, overlaid with $^{12}$CO\,(1--0) distribution in black contours plotted from 3$\sigma$ level. The white polygon in the figure shows the area observed by MOPRA, and the CO emission is localized toward our ALMA+APEX observation region, which is indicated as a green box. The 8\,$\mu$m PAH emission and H$\;${\sc i} show arc-like distribution extending from the northwest to the southeast, and the Giant Molecular Cloud (GMC) is located at the bottom of the arc. H$\;${\sc i} and PAH emission are extended further to the southwest from the bottom of the arc. H$\alpha$ is distributed in the northeast and south of the GMC. The total mass of the N113 GMC is estimated to be 4 $\times$ 10$^5$\,$M_{\odot}$ \citep{Fukui_2008} and is classified as a Type III GMC (actively forming stars and hosting young clusters), the most evolved of the three GMC types proposed by \citet{Kawamura_2009}. The GMC is associated with three young clusters, NGC~1874, NGC~1876, and NGC~1877 \citep{Bica_1992}, and the largest number of H$_2$O and OH masers, and the brightest H$_2$O maser in the entire LMC \citep[e.g., ][]{Green2008, Ellingsen2010}. There are about 130 Spitzer YSOs within a radius of 160 pc from the approximate center of N113 (RA, Dec) = (05$^\mathrm{h}$13$^\mathrm{m}$36$^\mathrm{s}$, $-$69\degr21\arcmin00\arcsec) \citep[e.g., ][]{Whitney2008, Gruendl2009, Carlson2012}. including 9 spectroscopically-confirmed massive YSOs, out of which 8 are associated with the brightest region in N113 \citep{, Seale_2009} 

The present paper is organized as follows. Section~\ref{sec:obs} describes the datasets used in the present work, and Section~\ref{sec:res} presents the results of the analysis of the ALMA+APEX $\twelvecoh$, $\thirteencoh$ and the ATCA H$\;${\sc i} data. Section~\ref{sec:dis} describes the distribution and kinematics of the H$\;${\sc i} gas, and discusses the origin of the GMC and the young stars. In Section 5, concluding remarks are given. 

\section{The Data} \label{sec:obs}
\subsection{ALMA Observations} \label{R:HCO_cont}

The N113 GMC was observed with ALMA 12\,m and 7\,m Arrays in Band 6 as part of project 2015.1.01388.S (PI M. Sewi{\l}o). In this paper, we use the data from the 
two narrow spectral windows covering the $^{12}$CO\,(2--1), $^{13}$CO\,(2--1), and C$^{18}$O\,(2--1) lines with the rest frequencies of 230.5380, 220.3987~GHz and 219.5603~GHz, respectively, and two $\sim$2~GHz windows centered on 231.7 and 216.9~GHz. N113 was observed with the 12\,m Array on March 10 and June 16, 2016, for a total of 13.1~min. per mosaic pointing over most of the map, on baselines from 15 to 704\,m.  It was also observed 19 times with the 7\,m Array between November 2 and December 17, 2015, for a total of 247 min. per mosaic pointing. The visibilities were calibrated and flagged with the ALMA data reduction pipeline version 2022.1.0.68 \citep{Hunter2023}.  Amplitude was calibrated using Uranus, J0538$-$4405, J0519$-$4546, and Callisto (different calibrators on different dates). The bandpass was calibrated using J0538$-$4405, J0522$-$3627, J0854$+$2006, J0006$-$0623, and time-varying gain using J0529$-$7245 (12\,m Array) and J0440$-$6952 (7\,m Array). The amplitude renormalization process was employed in the pipeline as has been standard for ALMA processing since 2021. The results for the two 2~GHz spectral windows centered on 231.7 and 216.9~GHz were reported in \citet{Sewilo_2018}. In this study, we analyze the $\twelvecoh$ and $\thirteencoh$ data, which effectively trace the molecular gas across the entire observed region.

The cube data were processed with version 6.5.6 of the Common Astronomy Software Applications (CASA; \citealt{CASA_2022}). The 12\,m array data were combined with the 7\,m array data in the visibility space. To enhance the flux recovery, we used the joint deconvolution scheme \texttt{sdintimaging}, where we input the single-dish data from the Atacama Pathfinder Experiment 12\,m telescope, APEX\footnote{This publication is based on data acquired with the Atacama Pathfinder EXperiment (APEX). APEX is a collaboration between the Max-Planck-Institut fur Radioastronomie, the European Southern Observatory, and the Onsala Space Observatory.} (see Section~\ref{sec:APEX}), as models and performed the CLEAN iteration to a threshold depth of 0.2 Jy\,beam$^{-1}$ with the Hogbom deconvolver applying the auto-masking technique \citep{Kepley2020}. Hereafter, we will refer to the ALMA+APEX images/observations/data products as ALMA images/observations/data products for simplicity.

The synthesized beam are 1\farcs07 $\times$ 0\farcs70 and 1\farcs17 $\times$ 0\farcs73 for the $\twelvecoh$ and the $\thirteencoh$, respectively. The typical spatial resolution is $\sim$0.2\,pc at the distance of LMC. The rms noise level in the $\twelvecoh$ and the $\thirteencoh$ data cubes is $\sim$0.5\,K for a velocity resolution of 0.2\,km\,s$^{-1}$.

\subsection{APEX Observations}\label{sec:APEX}

$^{12}$CO, $^{13}$CO, and C$^{18}$O (2--1) were observed with APEX between April 22 and July 11, 2019 as part of the combined Onsala Space Observatory (OSO) and 
Max-Planck-Institut für Radioastronomie (MPIfR) projects: O-0103.F-9315(A) (PI M. Sewi{\l}o) and M-0103.F-9515A (PI R. Chen), respectively. The observations were taken with the PI230 1~mm receiver with the spectral setup covering frequency ranges 213--221 GHz and 229--237 GHz simultaneously; this includes the three target lines of CO isotopes as well as lines of interests for serendipitous detections. The corresponding beam size for these frequency ranges is 26$\rlap.{''}9$--30$\rlap.{''}$0. The original spectral setup with the tuning frequency of 219.0 GHz had bad channels close to the C$^{18}$O line; the tuning frequency was then adjusted to 219.2 GHz and those affected C$^{18}$O data were excluded in the final data cube. 

A map covering the ALMA region was obtained using on-the-fly (OTF) mapping. Standard calibration was performed using R-Dor, Uranus, and o-Ceti. Data reduction was carried out using GILDAS/CLASS; to increase the signal-to-noise ratios in individual channels, contiguous channels were smoothed to a velocity resolution of 0.5 km~s$^{-1}$ and then baseline subtracted, resulting in the rms of $\sim$0.014, 0.013, and 0.010~K for $^{12}$CO, $^{13}$CO, and C$^{18}$O, respectively. The APEX data cubes were gridded in $9'' \times 9''$ ($\sim$3 pixels per beam) to facilitate comparisons with the ALMA data cubes. 


\subsection{Archival $\twelvecol$ Data} 
We used an archival $\twelvecol$ data taken with Mopra, the 22-m single-dish radio telescope located near Coonabarabran of Australia. The data was published by \cite{Wong_2011, Wong2017}. The spatial and velocity resolutions of the data cube used in the present paper are 45$''$ (11 pc at a distance of 50 kpc) and 0.5\,km\,s$^{-1}$, respectively. The rms noise level is 0.08~K. The $\twelvecol$ is used to show the large-scale CO distribution around N113 in the present paper. To show the consistency of spatial and velocity distribution in between this data to the APEX data, we compare the spatial distribution and velocity linewidth of this dataset to the APEX dataset in Appendix A.

\subsection{Archival H$\;${\sc i} Data} 
Archival data of the ATCA and Parkes 64~m telescope H$\;${\sc i} 21~cm line data \citep{Kim_2003} are used in the present study. The angular resolution of the combined ATCA and Parkes H$\;${\sc i} data is 60\arcsec ($\sim$15\,pc at the distance of LMC). The rms noise level of the data, determined from the line-free channels in the final H$\;${\sc i} data cube, is $\sim$2.4\,K for a velocity resolution of 1.649\,km\,s$^{-1}$.


All the three-dimensional data cubes used in this study are expressed in the $V_\mathrm{offset}$ velocity frame, which is defined as $V_{\mathrm{offset}} = V_{\mathrm{LSR}} - V_{\mathrm{rot}}$, where $V_{\mathrm{rot}}$ represents the rotation velocity of the galaxy modeled by \citet{Tsuge_2019}. Although $V_{\mathrm{rot}}$ varies position to position, we calculated a nominal value of $V_{\mathrm{rot}}$ to be 235~km\,s$^{-1}$ in the entire N113 region because the systematic difference of the model rotation velocity in our target field is less than 1~km s$^{-1}$. Therefore, we adopted
$V_{\mathrm{offset}} = V_{\mathrm{LSR}} - 235~\mathrm{km\,s}^{-1}$
for the following analysis.

\section{Results} \label{sec:res}

\subsection{The CO cloud distribution} \label{R:CO}
\begin{figure}[htb!]
    \centering
    \includegraphics[width=1.0\columnwidth]{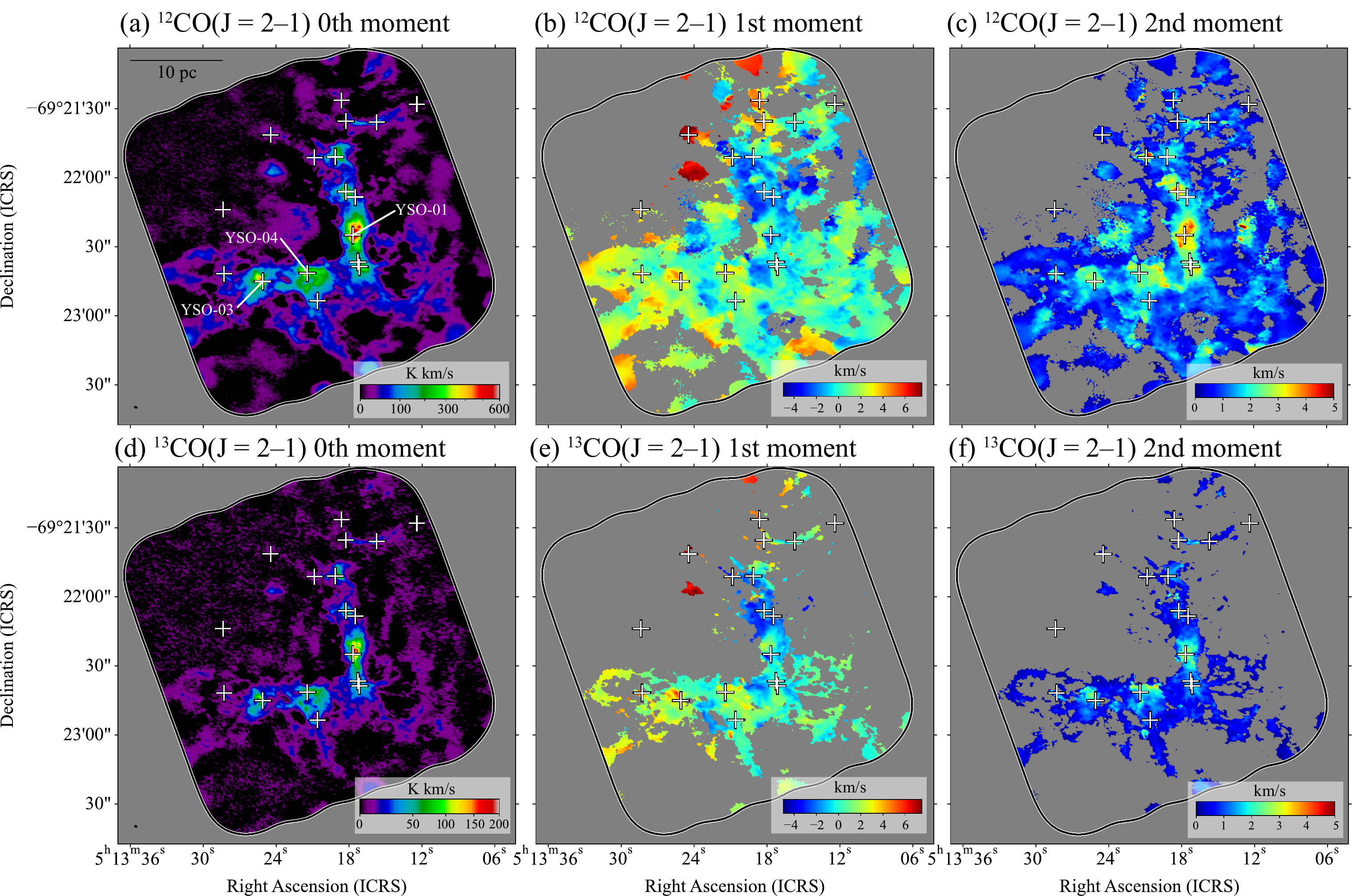}
    \caption{The $\twelvecoh$ (upper panel) and $\thirteencoh$ (lower panel) integrated intensity (moment 0; a and d), mean velocity (moment 1; b and c), and velocity dispersion (moment 2; c and f) images of N113. The white crosses represent the positions of YSOs and YSO candidates \citep{Whitney2008, Gruendl2009, Seale_2009, Carlson2012}. The beam size is shown in the lower left corner of each image.}
    \label{fig:COmom}
\end{figure}

\begin{figure}[htb!]
    \centering
    \includegraphics[width=0.7\columnwidth]{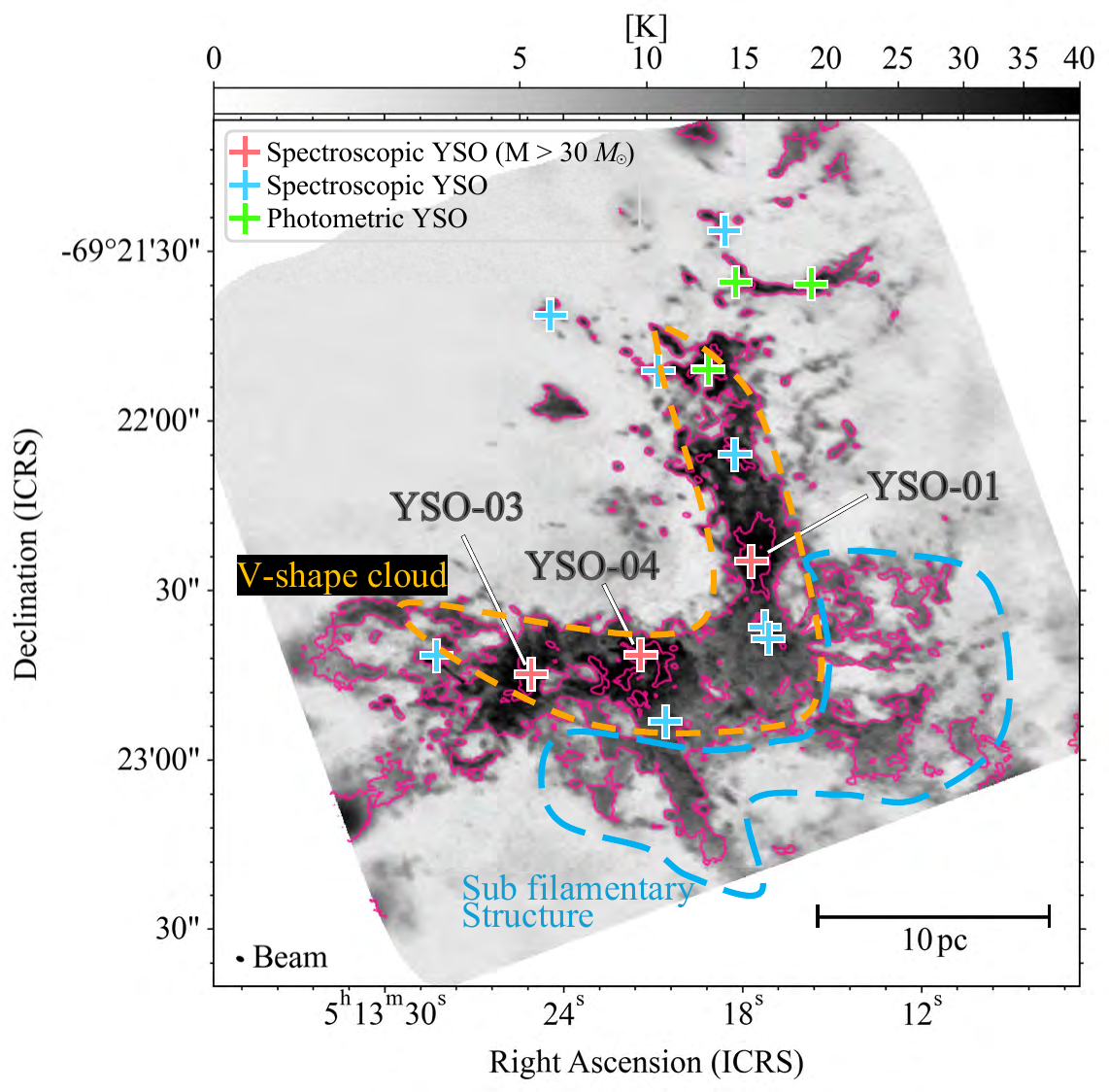}
    \caption{An ALMA+APEX resolved molecular gas view and prominent structures in the N113 region. The grayscale image is the peak brightness temperature map of $\twelvecoh$. The pink contours represent the $\thirteencoh$ peak temperature with contour levels of 3 ($\sim6\times$rms) and 15\,K. Pink and green crosses indicate massive YSOs \citep{Seale_2009, Sewilo_2010}, with those highlighted in pink with masses exceeding $\sim$30\,$M_{\odot}$ \citep{Ward_2016}. The cyan crosses indicate YSO candidates \citep{Gruendl2009, Carlson2012, Whitney2008}. The beam size is shown in the lower left. The bright V-shaped clouds of $^{12}$CO and $^{13}$CO and several sub-filamentary structures extending southwest and west are outlined with orange and cyan dashed lines, respectively.}
    \label{fig:struct}
\end{figure}

Figure~\ref{fig:COmom} shows the $\twelvecoh$ and $\thirteencoh$ integrated intensity (moment~0), the intensity weighted velocity (moment~1), and the velocity dispersion (moment~2) images of N113. Each moment is calculated in a velocity range from $-10$ to 10~km~s$^{-1}$ in the $V_\mathrm{offset}$ frame. The images reveal that both the $^{12}$CO and $^{13}$CO emission have two filamentary structures of $\sim$10--20\,pc in length elongated in the north-south and in the east-west direction, respectively. There are 14 YSOs in the region observed with ALMA+APEX \citep{Whitney2008,  Gruendl2009, Seale_2009, Carlson2012}, and they are associated with the CO gas, of which six YSOs are spectroscopically confirmed
\citep{Seale_2009} (See also Figure \ref{fig:1} and \ref{fig:Cloud_id}). Among them, YSO-01, YSO-03, and YSO-04 are associated with bright H$\alpha$ knots and have been reported to have masses greater than 30 $M_{\odot}$ \citep{Ward_2016}. The two filamentary structures form a $``$V-shape,$"$ with the vertex pointing toward a $\sim$50-degree position angle. The typical width of the $^{12}$CO filaments in N113, measured at one-third of the peak integrated intensity (around 600~K~km~s$^{-1}$), is 2--3 pc. This is about twice the width of the $^{13}$CO filaments, which typically have a 1~pc width at one-third of their peak integrated intensity (around 130~K~km~s$^{-1}$). A similar V-shaped, but fragmented structure has also been observed in ATCA observations of HCO$^+$ (1--0) and HCN (1--0) with a beam width of 6$''$ \citep{Seale_2012}. Enhanced by ALMA's high dynamic range observations, more continuous complex structures have been revealed. Specifically notable are several sub-filamentary structures (see Figure \ref{fig:struct}) extending southwest and west from the vertex of the V-shape and connected to the massive YSOs. In contrast, such sub-filamentary structures are not observed toward the opposite side, i.e., within the interior of the V-shape structure. This directional trait with sub-filaments emanating from the large scale V-shaped structure is also observed in the LMC-N159 region \citep{Tokuda_2022N159}. Further discussion on these features is presented in Section~\ref{sec:dis}.

The integrated intensity of $^{12}$CO exceeds 400\,K\,km\,s$^{-1}$ in N113 (see Figure~\ref{fig:COmom}(a)), particularly in the direction of massive protostars, which surpasses or matches the intensity observed near the Orion KL region in the Milky Way with similar resolution \citep{Nishimura_2015}. To estimate the total mass of the CO cloud observed with ALMA+APEX, we adopted an $X_{\rm CO}$ factor of  6.5 $\times$10$^{20}$\,cm$^{-2}$\,(K\,km\,s$^{-1}$)$^{-1}$ \citep{Naslim2018} because their estimate is based on CO datasets having closer resolution ($\sim$~0.7 pc) to our data ($\sim$~0.2 pc). Applying this $X_{\rm CO}$ factor and assuming a $\twelvecoh$/$\twelvecol$ ratio of 0.8 \citep{Nishimura_2015}, 
we calculated the total integrated intensity within the entire ALMA+APEX-observed region, resulting in a molecular gas mass of 2.0 $\times$ 10$^5$\,$M_{\odot}$. The calculation was performed in a velocity range from $-10$ to 10~km~s$^{-1}$ in the $V_\mathrm{offset}$ frame. The moment~1 maps in Figure~\ref{fig:COmom}(b,e) display a highly complex structure where the gas north of the vertex of the V-shape is blue-shifted and the gas east of the vertex is red-shifted. The difference in the cloud velocities across N113 is about 7~km s$^{-1}$. The velocities of sub-filaments are generally distributed around the central velocity. The main V-shaped structure shows variations in the north and northwest, with the northern part notably blueshifted by a few km\,s$^{-1}$. In regions hosting YSOs with masses of $\geqq$30\,$M_{\odot}$, the velocity dispersion exceeds 3\,km\,s$^{-1}$, which is larger than those in the rest of the cloud (1--2\,km\,s$^{-1}$). Specifically, the region around YSO-01 shows a velocity dispersion larger than 4\,km\,s$^{-1}$.


Figure~\ref{fig:struct} shows a summary of the CO emission features in N113 detected with ALMA+APEX, and their association with massive YSOs. The most intense filamentary molecular cloud forms the V-shaped structure. The YSOs are distributed along this V-shaped cloud, with the particularly massive and bright protostars, those with masses $\geqq$30\,$M_{\odot}$, located in regions of high column density and broader molecular lines. Weaker sub-filamentary components extend to the southwest and west from the vertex of the V-shaped structure and some of the massive YSOs. These structures and star formation characteristics are compared with a large-scale H$\;${\sc i} gas distribution in Sections~\ref{R:twoHI} and \ref{R:HI_disp}, and the inferred star formation scenario is discussed in Section~\ref{sec:dis}.

\subsection{Physical parameters of molecular clouds} \label{R:cloud_id}

To characterize cloud properties traced by $^{13}$CO, we examined the substructures within the cloud and their dynamical states. We used the \texttt{astrodendro} package \citep{Rosolowsky2008} to hierarchically segment and identify the cloud sub-structures as trunk, branch and, leaf. The method has been employed in several ALMA molecular cloud studies in the Magellanic Clouds \cite[e.g.,][]{Wong_2019,Wong_2022,Ohno_2023}. The algorithm requires three parameters to identify the significant entities: a minimum intensity (\texttt{min\_value}), a minimum number of voxels (\texttt{min\_pix}), and a minimum intensity difference between adjacent structures (\texttt{min\_delta}). We adopt \texttt{min\_value} of 2.5\,K, corresponding to 5 times the rms noise level. The parameter \texttt{min\_pix} is set to 50, requiring the identified structures to span at least 2 velocity channels or beam elements, where the spatial and velocity grids are 0\farcs18 and 0.2\,km\,s$^{-1}$, respectively. Finally, \texttt{min\_delta} is 2.5\,K, corresponding to 5 times the noise level. In this study, we focus on leaves, the smallest components of hierarchical structures with an effective radius of $\sim$0.2--2, to understand the characteristics of the denser $^{13}$CO gas, which is likely closer to prestellar and protostellar cores. The outputs of \texttt{astrodendro} used in the present work are cloud effective radius defined by $R_e$ = $\sqrt{\frac{A}{\pi}}$ where $A$ represents the area of the cloud, FWHM linewidth, and the total intensity, as well as the coordinates (R.A., Dec.) of the cloud center.

\begin{figure}[htb!]
    \centering
    \includegraphics[width=1.0\columnwidth]{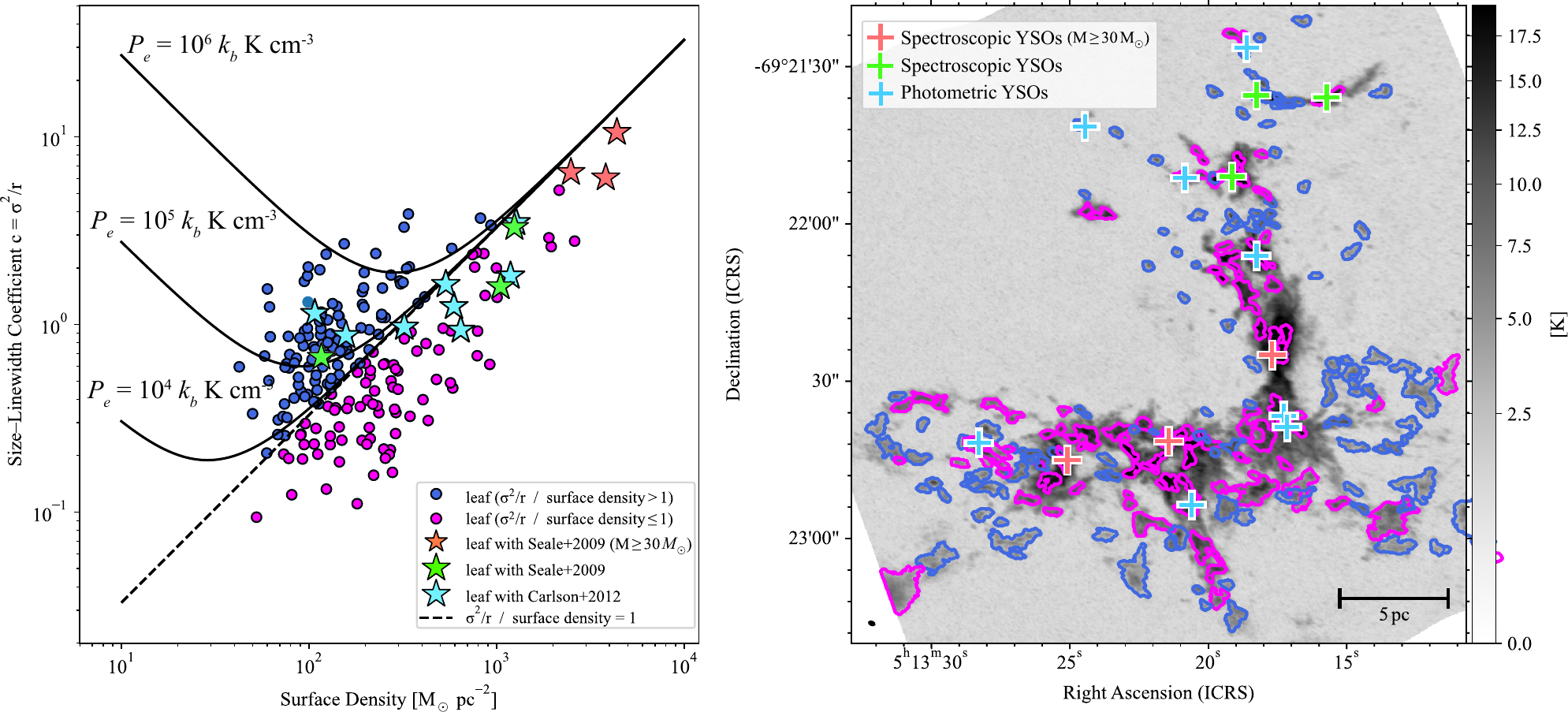}
    \caption{ {\bf Left panel}: Comparison of gas surface density and size-linewidth coefficients ($\sigma_v^2 / R$) for $^{13}$CO leaf structures (clumps) identified using \texttt{astrodendro}. Star symbols indicate leaves hosting previously-identified YSOs. Among these, leaves with spectroscopically determined YSOs are shown in red if the mass of the YSO exceeds 30 $M_{\odot}$, and in green if it is less than 30 $M_{\odot}$. Samples without spectroscopic confirmation, for which only photometric measurements are available (YSO candidates), are shown in cyan.
    The black dashed line shows the relation between the surface density and the size--linewidth coefficient for virial equilibrium in the absence of external pressure. 
    The black curves show the relations between the surface density and the size--linewidth coefficient for virial equilibrium under external pressures of 
    $P_{\rm ext} = 10^{4}\,k_{\rm B}\,{\rm K\,cm^{-3}}$, 
    $10^{5}\,k_{\rm B}\,{\rm K\,cm^{-3}}$, and 
    $10^{6}\,k_{\rm B}\,{\rm K\,cm^{-3}}$, where $k_{\rm B}$ is the Boltzmann constant. 
    {\bf Right panel}: The grayscale image is the peak intensity map of $\thirteencoh$. Magenta and blue contours represent leaf structures in virial equilibrium with and without external pressure, respectively.
    Crosses are the same as in Figure \ref{fig:struct}}
    \label{fig:Cloud_id}
\end{figure}

To investigate whether the molecular clouds are gravitationally bound, we examined the relationship between their surface density, size, and velocity dispersion (Figure \ref{fig:Cloud_id}). To estimate the surface density, we assumed local thermodynamic equilibrium (LTE) to calculate the CO surface density. Then, we adopted [H$_2$]/[$^{13}$CO] abundance ratio of 3.2 $\times$ 10$^{6}$, which is commonly used in early works in the LMC \citep[e.g.,][]{Heikkila_1999,Mizuno_2010} to convert the CO surface density to H$_2$ surface density. We constructed a diagnostic diagram following \citet{Heyer2009}, where the x-axis represents the surface density and the y-axis represents the size--linewidth coefficient, $\sigma_v^2 / R$. About half of the molecular clumps appear to be gravitationally bound, while the other half appear to be supported by external pressure if we assume that they are not transient structures. All CO clumps associated with YSOs, except for three, are gravitationally bound. A notable feature is that gravitationally bound clumps are distributed along the V-shaped cloud, while clumps in other locations are less bound. 

As shown in the left panel, the clumps highlighted in a coral-pink color which hold the YSOs with masses exceeding $\sim$30\,$M_{\odot}$, belong to the most massive category, on the order of 10$^{3}$\,$M_{\odot}$. The leaves associated with other spectroscopic YSOs, indicated in green and located on the northern side, are less massive; one of these has a relatively higher size-linewidth coefficient.

\subsection{H$\;${\sc i} clouds toward N113} \label{R:twoHI}
\begin{figure}[htb!]
    \centering
    \includegraphics[width=\columnwidth]{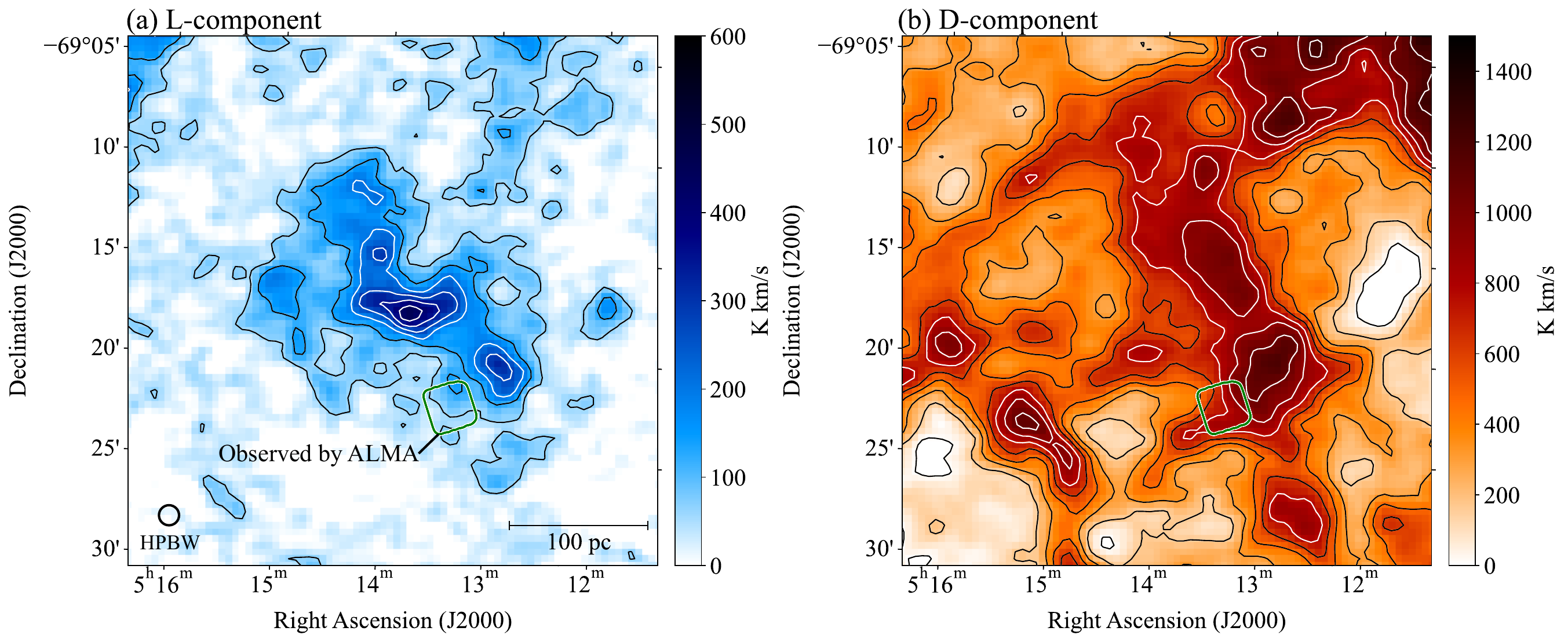}
    \caption{Integrated intensity maps of H$\;${\sc i} clouds around N113. The velocity range over which the emission was integrated is ($-50.1$, $-30.5$)\,km\,s$^{-1}$ for the L-component (panel a) and ($-10.4$, 9.8)\,km\,s$^{-1}$ for the D-component (panel b). The green square indicates the ALMA+APEX FoV. Contours are plotted every 70 and 150 K km~s$^{-1}$ from 60 and 10 K km~s$^{-1}$ for panels a and b, respectively.}
    \label{fig:twoHI}
\end{figure}

Figures~\ref{fig:twoHI}(a) and (b) show the H$\;${\sc i} distribution in the ($-50.1$, $-30.5$)\,km\,s$^{-1}$ and ($-10.4$, $-9.8$)\,km\,s$^{-1}$ velocity ranges in the area around N113 that is nearly ten times larger than the ALMA+APEX field. The two velocity ranges correspond to the low-velocity (L-)component and the disk (D-)component defined in earlier studies \citep[e.g,][]{Luks_1992,Fukui_2017,Tsuge_2019}. The L-component is particularly concentrated on the southeastern side of the LMC but is also locally scattered across the entire galaxy. A convincing interpretation of the L-component is that it consists of the disturbed and/or accreted gas by the last close encounter with the SMC 2 Gyr ago \citep{Fujimoto_1990,Bekki_2007, Tsuge_2024}. The D-component represents the disk component of the galaxy. Figure~\ref{fig:twoHI}(a) indicates that the L-component associated with N113 is elongated with a size of 200\,pc $\times$ 100\,pc and a position angle of about $-$45 degrees, and Figure~\ref{fig:twoHI}(b) indicates that the low intensity part of the D-component, which have the integrated intensity lower than 600 K km~s$^{-1}$, has a similar elongation with a position angle of about $-$30\,degrees.

\begin{figure}[htb!]
    \centering
    \includegraphics[width=0.43\columnwidth]{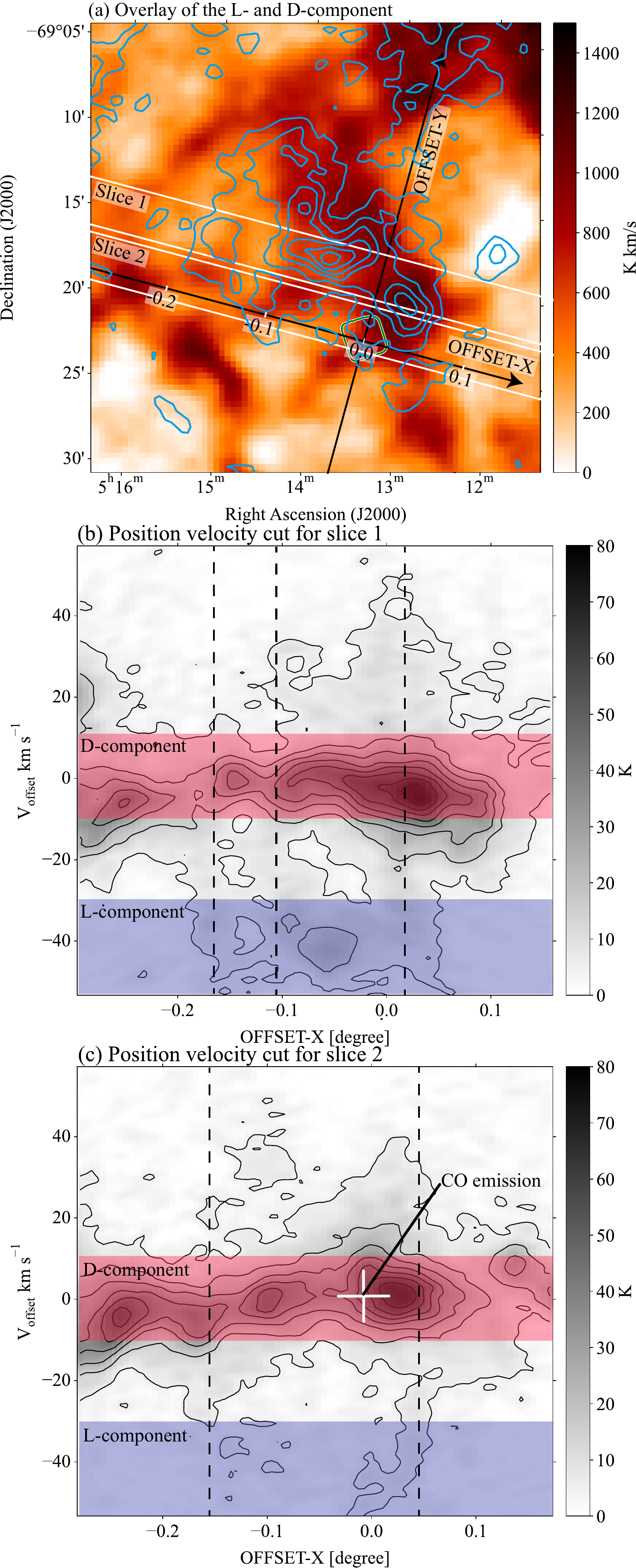}
    \caption{(a) Integrated intensity map of H$\;${\sc i} clouds around N113 (image: D-component, contours: L-component). The green square indicates the ALMA+APEX FoV. (b) H$\;${\sc i} position–velocity cut along `slice 1' indicated in (a). The diagram is along the OFFSET-X and OFFSET-Y coordinates, which are rotated 15 degrees clockwise from the J2000 coordinate, with a rotation origin being the centre of the ALMA+APEX FoV. Contours are plotted every 10 K starting at 5 K. (c) Same as (b), but for slice 2 in (a). The white cross indicates an approximate spatial and spectral extent of the CO emission in N113. The black dashed lines in the figure indicate the approximate offset-X positions where the I-component is clearly seen as a bridge connecting the L-component and the D-component.}
    \label{fig:HIpv}
\end{figure}

Figure~\ref{fig:HIpv}(a) shows an overlay of the two H$\;${\sc i} clouds. The N113 GMC, ten times smaller than these H$\;${\sc i} clouds, is located in the southeast of the southwestern tips of these H$\;${\sc i} clouds. Figures~\ref{fig:HIpv}(b) and (c) show two position-velocity diagrams in the `X offset' for two `Y offset' values close to the ALMA+APEX field as shown in Figure~\ref{fig:HIpv}(a). Figure~\ref{fig:HIpv}(b) shows that the L-component and the D-component are linked by intermediate-velocity bridge features, which are part of the I-component produced by the interaction between L- and D- components \citep{Tsuge_2019}, at three positions of X offset=$-$0.18, $-$0.12 and 0.05 degrees. We find that the D-component has enhanced linewidth toward the bridge feature at X offset=$-$0.1 degrees. Similar features connecting the L- and D-components are found toward R136 \citep{Fukui_2017} and N159 (\citealt{Fukui_2019}; see their Figure 7a), and are believed to be due to collisional interactions. We suggest that the bridge features represent dynamical interactions between the two clouds as demonstrated by numerical simulations of CCCs \citep{Takahira_2014}.

\begin{figure}[htb!]
    \centering
    \includegraphics[width=1.0\columnwidth]{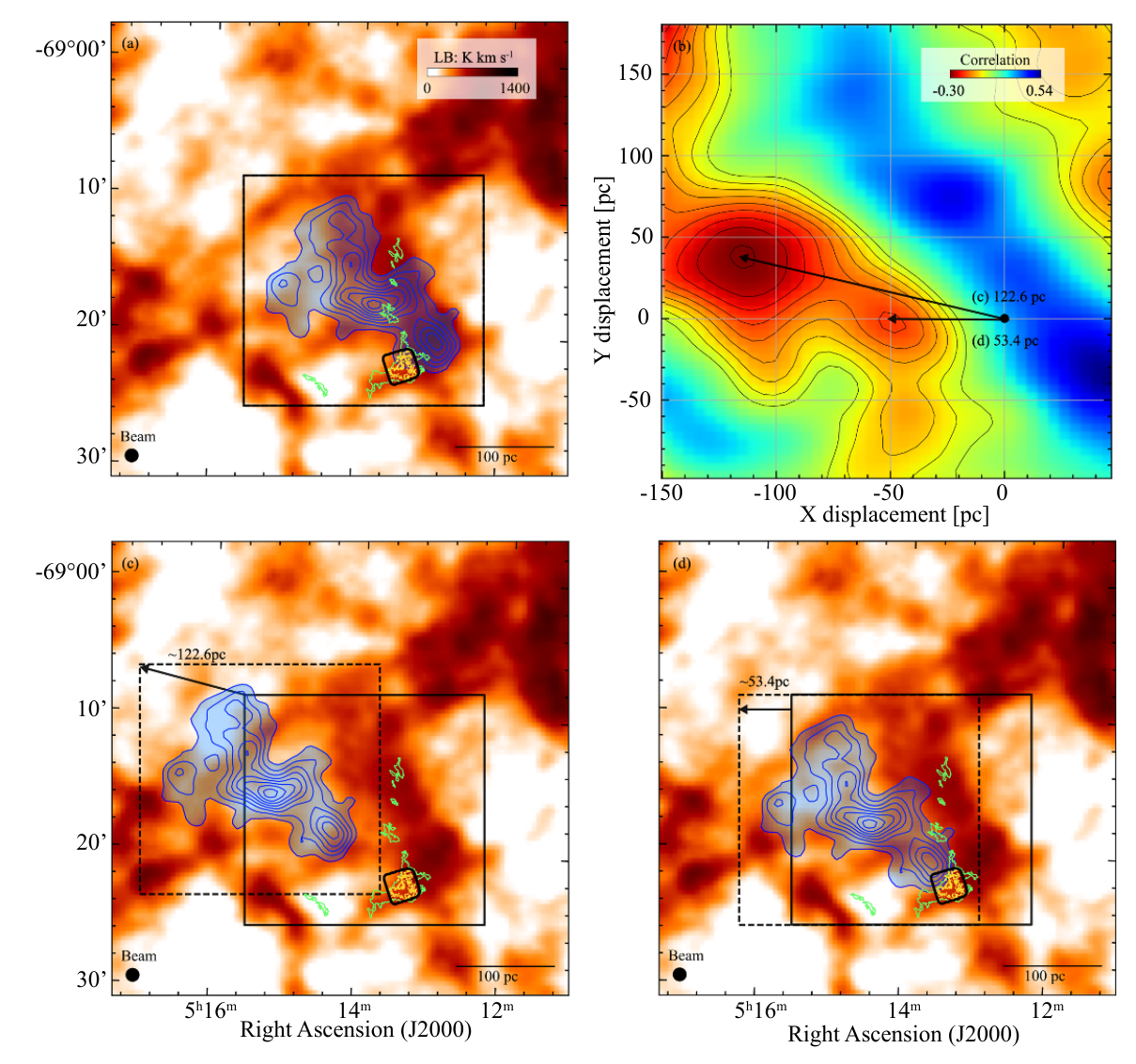}
    \caption{(a) H$\;${\sc i} integrated intensity map of the D-component with a velocity range of ($-$10, 10)\,km\,s$^{-1}$ with the L-component with a velocity range of ($-$50, $-$30)\,km\,s$^{-1}$ contours overlaid. (b) The spatial distribution of the correlation coefficient was derived by repeatedly calculating the correlation between the integrated intensity maps of the L-component and the D-component while systematically shifting one map pixel by pixel along the X direction (parallel to RA) and the Y direction (parallel to DEC). The lower value of the correlation coefficients indicates the regions with spatial anti-correlation, where the two components have complementary distributions. Two local minima can be seen (see the text). (c) and (d) Same image as (a), but with the L-component contours displaced. The solid box indicates the initial position of the L-component, and the dashed box indicates the displaced positions. The yellow and green contours correspond to the ALMA+APEX $\twelvecoh$ and the Spitzer/IRAC 8 $\mu$m emission with contour levels of 3~$\sigma$ and 5 MJy~sr$^{-1}$, respectively.} 
    \label{fig:HIdisp}
\end{figure}

 We calculated the column densities and masses of the H$\;${\sc i} clouds as follows. The column density of H$\;${\sc i}, $N$({\rm H\;{\sc i}}), is calculated from the radiative transfer equation for each pixel by assuming the optically thin condition, using the equation of 
$N$(H$\;${\sc i}) (cm$^{-2}$) = 1.823 $\times$ 10$^{18}$~$W$(H$\;${\sc i}) (K\,km\,s$^{-1}$) \citep{Dickey1990}, 
where $W$(H$\;${\sc i}) is the integrated intensity of the H$\;${\sc i} emission line, given as $W$(H$\;${\sc i}) = $\int$ $T_{\rm b}dv$ with $T_{\rm b}$ representing the observed H$\;${\sc i} brightness temperature. The peak column density is $N$({\rm H\;{\sc i}}) = 4.4 $\times$ 10$^{21}$\,cm$^{-2}$ and the total H$\;${\sc i} mass is 9 $\times$ 10$^6$\,$M_{\odot}$ in the entire area shown in Figure~\ref{fig:twoHI} with a velocity range from $-50$~km~s$^{-1}$ to 10~km~s$^{-1}$. The L-component has a peak column density of $N$(H$\;${\sc i}) = 8.3 $\times$ 10$^{20}$\,cm$^{-2}$ at (RA, Dec)=(05$^\mathrm{h}$13$^\mathrm{m}$46$^\mathrm{s}$, $-69$\degr17\arcmin25\arcsec). The total H$\;${\sc i} mass of the L-component is 4 $\times$ 10$^5$\,$M_{\odot}$. The D-component has a peak column density of $N$(H$\;${\sc i}) = 2.2 $\times$ 10$^{21}$\,cm$^{-2}$ at (RA, Dec) = ($05^\mathrm{h}12^\mathrm{m}52^\mathrm{s}$, $-69$\degr19\arcmin56\arcsec) and total mass of the D-component is 5 $\times$ 10$^6$\,$M_{\odot}$ in an area shown in Figure~\ref{fig:twoHI}. We note that the H$\;${\sc i} column densities given above are likely lower than those at smaller scales due to dilution by the low resolution ($\sim$10 pc) of the H I observations.

\begin{figure}[htb!]
    \centering
    \includegraphics[width=1.0\columnwidth]{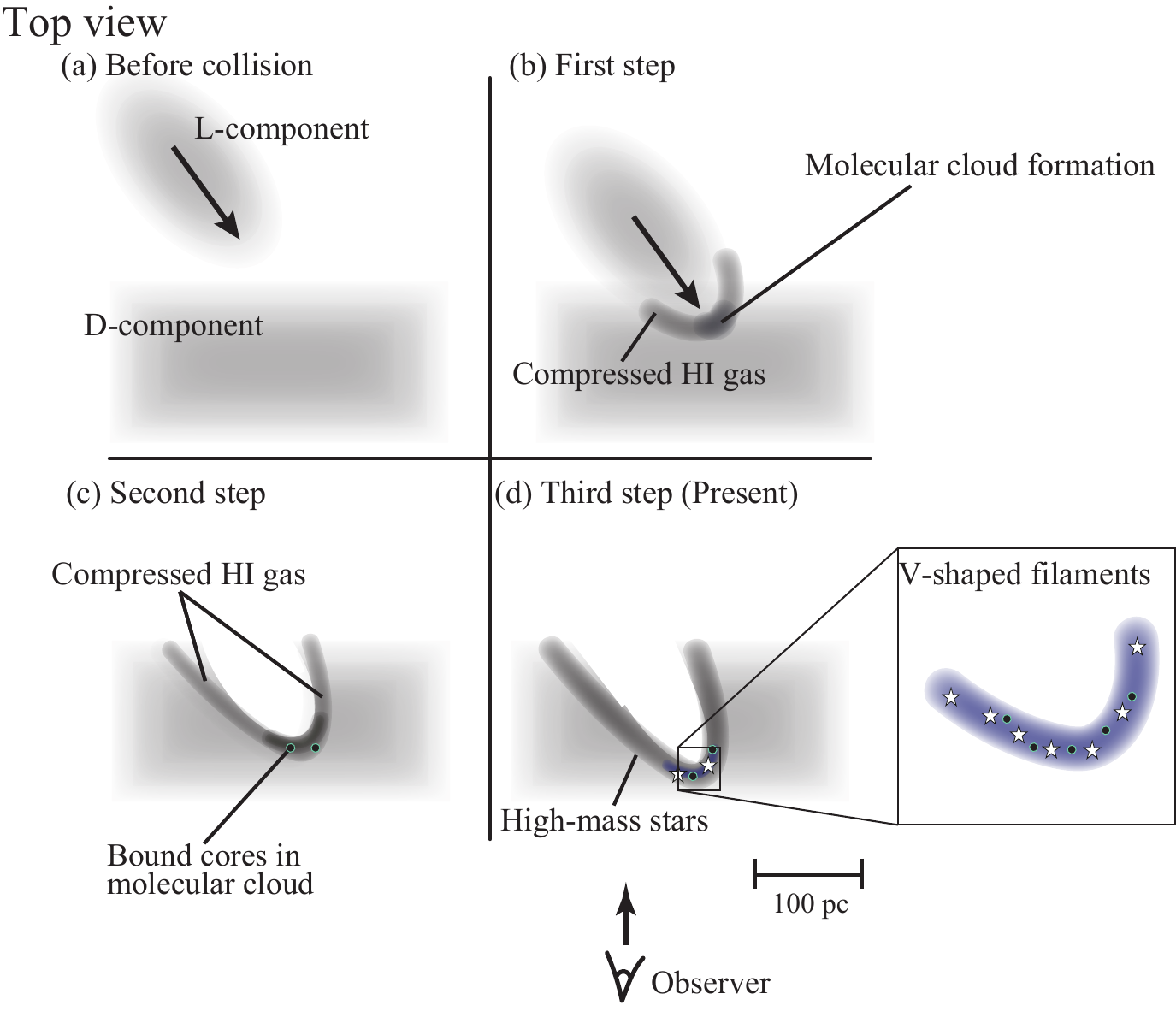}
    \caption{The schematic top view of the cloud-cloud collision in the N113 region.}
    \label{fig:schematic}
\end{figure}

\section{DISCUSSION} \label{sec:dis}
\subsection{Evidence of a CCC between the two clouds in N113}\label{R:HI_disp}
The typical observational signatures of CCCs are: (1) a bridge feature connecting the two velocity components in velocity space, and (2) a complementary (spatially anti-correlated) distribution between the two clouds, often accompanied by (3) a U-shaped morphology in the larger cloud, as reviewed by \citet{Fukui_2021}. These signatures are characteristic of collisions between a large and a small cloud, as such interactions are presumed to be more common than collisions between clouds of similar size and density.

These three CCC signatures are the consequence of the dynamical interactions between the two clouds. The bridge feature arises from momentum exchange between the two clouds, which produces intermediate-velocity gas located between the red-shifted and blue-shifted components in the velocity space. The complementary distribution results from the small cloud penetrating deeply into the larger cloud and merging with the gas of the larger cloud. The gas in the larger cloud that lies along the trajectory of the small cloud becomes entrained and/or incorporated, creating a cavity in the larger cloud that mirrors the shape of the smaller cloud. This leads to the observed complementary (spatial anti-correlation) between the two clouds. When the collision direction is significantly inclined with respect to the line of sight, the complementary distribution appears displaced, as discussed in detail by \citet{Fukui_2018}. The U-shaped morphology is likewise a consequence of gas merging, wherein the larger cloud is hollowed out in a U-shape due to the inclined geometry of the collision. Note that the three features were primarily modeled for molecular clouds \citep[e.g., ][]{Fukui_2018}; however, since the essential physics involves the collision of gas clouds with different sizes and densities, it is highly likely that the same features can also be observed in collisions between atomic clouds. Indeed, such CCC signatures have been observed in other regions of the LMC \citep{Fukui_2017, Tsuge_2019, Tsuge_2024}.


The gas properties in N113 are consistent with the three signatures of CCC. The L- and D-components show the clear bridge feature, as indicated by the dashed lines on the PV diagrams (Figure \ref{fig:HIpv}bc), and the D-component shows the U-shaped morphology (Figure \ref{fig:twoHI}b). The D-component has an intensity depression as indicated in the ellipse in Figure \ref{fig:twoHI}b. The depression shows a roughly elliptical shape and seems to fit the shape of the L-component in Figure \ref{fig:twoHI}a, indicating the possible complementary distribution.

To evaluate the complementarity and displacement between the two atomic clouds, we calculated the Pearson's correlation coefficients of their emissions using the method developed by \citet{Fujita_2021}. In order to obtain the spatial distribution of the Pearson's correlation coefficients, we shifted the integrated intensity map of the L-component pixel-by-pixel and calculated the Pearson's correlation coefficients between D- and L-components at each offset pixel. Then, we mapped the plane-of-the-sky distribution of the correlation coefficients, as shown in Figure \ref{fig:HIdisp}b. We see two minima of the coefficients, which indicate possible complementary distributions between the two clouds, a signature of CCC, and applied displacements corresponding to the two minima in Figures~\ref{fig:HIdisp}(c) and (d).  Figure~\ref{fig:HIdisp}(c) shows a case for a displacement of 122.6\,pc, the minimum of the coefficients, and Figure~\ref{fig:HIdisp}(d) a case for a displacement of 53.4\,pc, the second minimum of the coefficients. Although the former case has a higher coefficient than the latter, the complementary matching between the L-component with the GMC and the 8\,$\mu$m features at the tip of the L-component is more reasonable as shown by an overlay in Figure~\ref{fig:HIdisp}(d). We, therefore, adopt the latter displacement in the following as a probable signature of a CCC. Our interpretation is that the L-component collided with the D-component in the past and triggered the formation of the GMC and the 8\,$\mu$m PAH features by the collisional compression. The collision time scale is estimated to be 53.4\,pc/40\,km\,s$^{-1}$ $\sim$1.3 Myr for an assumed angle of 45 degrees between the line of sight and the collision path in 3D.

The displacement in the complementary distribution indicates the large angle difference between the cloud motion vector and the line-of-sight, and hence we can estimate the cloud motion by the length of the displacement and radial velocity. As the L-component fits well with the D-component's cavity by applying the displacement of 53.4 pc toward the east, the L-component has traveled from the west of the D-component. In addition, the L-component has a blue-shifted velocity, indicating that the L-component is coming toward us. Hence, we can presume that the L-component has moved from the far eastern side of the D-component, collided with the D-component to create a cavity, and then passed through the D-component, such that it is now located on the near side relative to the D-component, as shown in the schematic picture in Figure \ref{fig:schematic}. 

This three-dimensional motion is not only consistent with the H$\;${\sc i} gas distribution of the red and L-components but also agrees with the $\twelvecol$ and 8~$\mu$m distribution. According to the wider field observations by \citet[Mopra, HPBW$\sim$45$''$]{Wong_2011}, the CO cloud in this region is localized within the field observed by ALMA+APEX (see appendix \ref{sec:appA}), and 8 $\mu$m PAH emission is also localized toward the CO clouds as shown in Figure \ref{fig:1}. The resolved by ALMA+APEX V-shaped CO clouds with an open angle of 10--20 degrees can be explained by assuming that local shock compression played a role in the formation of N113 GMC, supporting the scheme of molecular cloud formation via collision. 

\subsection{The CCC scenario in N113}
Based on the three-dimensional cloud motion discussed above, we propose a scenario in which a CCC took place in three steps: (step 1) the formation of the H I cavity, followed by (step 2) the formation of a V-shaped giant molecular cloud, and (step 3) the high-mass star formation. The details of each step are described in the following.

\textit{First step}: CCC was initiated by the collision between the two H$\;${\sc i} clouds and continued until 1.3\,Myr ago (see below), during which the L-component collided with the D-component and created the complementary distribution and the two bridge velocity features at the eastern and western ends of the L-component between them. If we assume that in the process, the L-component traveled over 50\,pc, the duration of the first step is estimated to be $\sim$1\,Myr for the projected velocity difference of 40\,km\,s$^{-1}$. The collision of the L-component likely occurred from the far side of the LMC, as in the case of the H$\;${\sc i} Ridge \citep{Fukui_2017}. The projected collision velocity is 40\,km\,s$^{-1}$, but can be 56\,km\,s$^{-1}$ if 45 degrees is assumed as an angle between the collision velocity and the sight line. According to the displacement derived in Section~\ref{R:HI_disp}, the epoch when the L-component reached the position 53 pc to the east from the present position (Figure~\ref{fig:HIdisp}(d)) is estimated to be 1.3\,Myr ago, and the collision path is nearly in the east-west direction. 

\textit{Second step}: This step of the CCC began 1.3\,Myr ago when the L-component reached the northwestern cavity wall of the D-component (Figure~\ref{fig:HIdisp}(d)). We hypothesize that, at the time, an H$\;${\sc i} spherical cloud of a 100-pc diameter, part of the blue-shifted H$\;${\sc i} cloud, moved to the west and collided with the D-component. This hypothesis is motivated by the previous modeling of the fan-shaped CO clouds in N159E and W \citep{Fukui_2019,Tokuda_2019,Tokuda_2022N159}, which were formed by a similar incident cloud according to the MHD model by \citet{Inoue_2018}. The presence of sub-filaments extending west and southwest from the main V-shaped cloud (Figures~\ref{fig:struct} and \ref{fig:COmom}) indicates collisions in that direction. As detailed in \cite{Tokuda_2022N159} and their Figure 8, this is evidence that the relatively high-column-density subpeaks created by turbulent perturbations existing before the collision could not decelerate completely and partially penetrated the compressed V-shaped layer. In the case of N113, the lack of significant velocity difference between the main cloud and the sub-filament (Figure~\ref{fig:COmom}(b, e)) suggests that the collision occurred perpendicularly to the observer's line of sight.

In order to further examine the second step, we compare the present results to the MHD simulation by \citet{Inoue_2018}, which assumes that a spherical cloud causes a head-on collision with an extended cloud having plane-like surface. In the collision the magnetic field is bent into a V-shape by the mechanism of \citet[see their Figure 1]{Inoue_2013}. The compressed interface gas between the two clouds flows along the field line, forming a dense, cone-shaped wall with density higher than 10$^{4}$\,cm$^{-3}$, where the initial density is around 10$^3$ cm$^{-3}$ prior to the collision (see \citealt{Fukui_2021rapid} for the detailed density distribution). The cone wall is thin and consists of several filamentary gas structures with a V-shaped cross-section around the cone's symmetry axis. The cone's height is roughly determined by the product of the velocity along the cone axis and the duration of the collision. In this case, the velocity is 7\,km\,s$^{-1}$, calculated as 10\,pc divided by the assumed collision duration of 1.3\,Myr. This velocity is much smaller than the initial velocity of 40\,km\,s$^{-1}$, indicating that the incident cloud was significantly decelerated from 56\,km\,s$^{-1}$ in the second step. This is plausible if the colliding L-component has about ten times lower density than the D-component. The density and the timescale provide conditions which can form CO gas in the compressed layer, where the H$_2$ formation timescale given as 10$^7$($n$/100\,cm$^{-3}$)$^{-1}$ yr \citep{Hollenbach_1971} becomes $\sim$1\,Myr for number density ($n$) of 10$^3$\,cm$^{-3}$.  High-mass stars are being formed in the dense CO clumps whose density can reach more than 10$^{5}$\,cm$^{-3}$ \citep{Inoue_2013,Fukui_2021rapid} and are observed as massive YSOs (Figures~\ref{fig:COmom} and \ref{fig:struct}).

The CCC model of \cite{Inoue_2018} provides a cone-shaped, hub-filament formation scenario for the high-mass stars in the N159 clouds \citep{Fukui_2019,Tokuda_2019,Tokuda_2022N159} and other LMC systems \citep{Tokuda_2023LMC,Sewilo_2023}. Further quantitative follow-up numerical model (Maity et al. in prep.) and the observational data analysis of the N159E and 30~Dor clouds are ongoing (Yamada et al. in prep; Fukui et al. in prep.). The heights of the cones with vertices ranging from $\sim$3 to $\sim$40 pc and the open angle ranging from $\sim$10 to $\sim$40 deg. The largest cone-type filament in R136 is due to the high collision velocity exceeding 50\,km\,s$^{-1}$ and the large diameter of the incident cloud ($\sim$40\,pc). On the other hand, the typical height and bottom diameter of the cone in N159E are 3\,pc and 2\,pc, respectively, for a collision velocity of a few $\times$10\,km\,s$^{-1}$. The size of the fan-shape (or simply, V-shape) cloud in N113 is 10\,pc, which lies between the values for N159E and R136.

\textit{Third step}: Following the formation of the V-shaped cloud, the star formation commenced in the dense cores along the V-shape cloud. A notable characteristics of the CCCs are the clustering of star formation\citep[e.g., ][]{Yamada2021}. The star formation in N113 is distributed over the cloud with a typical separation of $\sim$5\,pc and there is no massive YSO with a mass of $\gtrsim$30\,$M_{\odot}$ around the cone vertex. In the N159 clouds, the highest mass YSOs are concentrated near the vertex of the cone, which has the highest density \citep{Fukui_2019,Tokuda_2019,Tokuda_2022N159}. 
Generally, the density distribution of the fan in N159 evolves such that the compressed gas flows toward the vertex. In contrast, in N113, the compressed gas likely becomes self-gravitating soon after compression in the cone wall without flowing over a large distance. This is probably caused by the high column density of the compressed gas layer in N113, as evidenced by the presence of six massive stars along the V-shaped structure and many gravitationally bound substructures (Figure~\ref{fig:Cloud_id}). Among these structures, the more massive blobs are likely to collapse toward high-mass star formation eventually. In such cases, the spatial alignment of O-type stars in the eventually formed cluster is likely to maintain a V-shaped structure at a certain level. A V-shaped distribution of stars is also observed in the supernova remnant 30~Dor~C in the LMC (see Figure~1(a) in \citealt{Yamane_2021}). On the other hand, the gas in the N159 clouds seems to flow more ballistically than that in N113, and is less affected by the self-gravity. Although the initial gas relative velocity between L- and D-components before the deceleration due to the collision is about 100\,km\,s$^{-1}$ in both N159 and N113 regions, the sizes of the V-shaped structures are 10--20 pc in N113 and 1--5 pc in N159. This may imply that the colliding gas in the N159 regions appears to have denser and more compact structures that form a ballistic flow, unlike that in N113. As a result, the flow penetrated to the vertex before self-gravity could dominate the region.

The N113 GMC has two filamentary clouds of 10~pc length and the average H$_2$ column density of $\sim$5 $\times$10$^{22}$\,cm$^{-2}$. On the other hand, the R136 cloud has several filamentary clouds of 40\,pc length and the average H$_2$ column density is $\sim$3 $\times$10$^{22}$\,cm$^{-2}$, lower than that in N113. 
Star formation in N113 likely occurs in the V-shaped filamentary clouds, while in the R136 cloud, it predominantly happens in the dense condensation outside the cone wall \citep{Nayak_2023}. This difference in star formation location is possibly due to the varying collision velocities: $\sim$7\,km\,s$^{-1}$ in N113 and 70\,km\,s$^{-1}$ in R136. The higher velocity creates a longer cone with a lower column density in the cone wall, while the lower velocity results in a smaller cone and higher column density in the filamentary clouds. The difference in collision velocity likely regulates star formation, although other factors may also contribute.

Further, we pay attention to the difference of the projected distribution of the clouds between R136 and N113. 
N113 has only two filamentary clouds, while R136 contains several. This variation may be due to differences in the initial gas distribution in the D-component. In the case of R136, the model analysis shows that the gas distribution in the D-component is of a half-cone shape because the velocity of the CO gas corresponds to only a half-cone on the far side instead of a whole cone. This was suggested to be due to the density decrease of the near side of the D-component. 
In the case of N113, we suggest that the plane cloud in the model was not extended much vertically but was rather thin. As a result, the cone forms in 2D instead of 3D, explaining the V shape without a 3D cone. Consequently, the convergence of the colliding gas into a small volume is less efficient than in R136.

\section{CONCLUDING REMARKS}\label{sec:summary}
We have carried out the analysis of the ALMA+APEX $\twelvecoh$ and $\thirteencoh$ data for the H$\;${\sc ii} region N113 in the LMC at a resolution of 1\farcs1 $\times$ 0\farcs7 ($\sim$0.2~pc). While the high resolution image revealed detailed structures and physical properties of molecular gas in N113, we also utilized archival H$\;${\sc i} data ($\sim$15 pc resolution) to examine the formation mechanism of such structures and star formation inside in terms of 100~pc scale gas dynamics. The main conclusions of the present study are summarized below.

\begin{itemize}
\item[1.] The N113 region consists of the GMC with a total mass of 2.0 $\times$ 10$^5$\,$M_{\odot}$ based on the $\twelvecoh$ observations with the resolution of $\sim$0.2\,pc. The GMC is characterized by two prominent 10 pc filaments forming a V-shaped clouds accompanied by sub-filamentary structures toward the east and southeast. The CO cloud contains six embedded massive YSOs (three mass of $\sim$30\,$M_{\odot}$), along with several Wolf-Rayet (WR) and/or OB stars within a radius of a 100\,pc. We decomposed the $^{13}$CO emission into approximately 200 substructures (clumps) and estimated their size-linewidth coefficients ($\sigma_v^2 / R$) and surface density based on the LTE approximation. The clumps with higher surface density, indicating gravitationally bound sources, tend to be distributed along the prominent V-shaped structure and be associated with YSOs. In contrast, the clumps located on the outside tend to be less gravitationally bound.

\item[2.] A comparison with the ATCA H$\;${\sc i} data reveals that the N113 GMC is associated with H$\;${\sc i} clouds at two velocities separated by $\sim$40\,km\,s$^{-1}$. The H$\;${\sc i} gas mass of the L-(blue-shifted) and D-(red-shifted) components (extending over more than 100 pc) are $\sim$4 $\times$ 10$^{5}$\,$M_{\odot}$ and $\sim$9 $\times$ 10$^{6}$\,$M_{\odot}$, respectively. The redshifted cloud has a cavity with a diameter of 100\,pc, complemented by a blueshifted cloud. Displacing the blueshifted cloud approximately 50\,pc to the east further emphasizes this complementary distribution.
The specific spatial distribution, along with the bridge features linking the two clouds in velocity, provides strong evidence for the dynamical interaction between the two clouds in N113.

\item[3.] Based on our results, we propose the scenario that the formation of the N113 system ((the GMC and massive stars)) was triggered by the CCC in the region. The process consists of three steps: an H$\;${\sc i}  cloud collision, the formation of a large cavity in one of the clouds, and the formation of the GMC and massive stars The first step occurred more than 1.3\,Myr ago, followed by the second step to form the V-shaped molecular cloud with gravitationally bound substructures, which collapsed into high-mass stars as the third step. These steps are supported by previous papers on the CCC picture modeled by MHD numerical simulations. The observational results reported in  N159 and R136 in the LMC also agree with the triggered formation of the high-mass stars and clusters via the three steps. This study further emphasize the importance of lar scale H$\;${\sc i} collision driven by LMC-SMC interaction as a trigger of GMC formation and high-mass star formation.

\end{itemize}

\begin{acknowledgments}

This paper makes use of the following ALMA data: ADS/JAO.ALMA\#2015.1.01388.S. ALMA is a partnership of ESO (representing its member states), the NSF (USA), and NINS (Japan), together with the NRC (Canada), MOST and ASIAA (Taiwan), and KASI (Republic of Korea), in cooperation with the Republic of Chile. The Joint ALMA Observatory is operated by the ESO, AUI/NRAO, and NAOJ. Cerro Tololo Inter-American Observatory (CTIO) is operated by the Association of Universities for Research in Astronomy Inc. (AURA), under a cooperative agreement with the National Science Foundation (NSF) as part of the National Optical Astronomy Observatories (NOAO). The MCELS is funded through the support of the Dean B. McLaughlin fund at the University of Michigan and through NSF grant 9540747. This work was supported by grants-in-aid for scientific research (KAKENHI) of the Japan Society for the Promotion of Science (JSPS; grant Nos. JP20H01945, JP20H05645, JP21K13962, JP22H00152, JP23H00129, 22KJ1604, and 25K23399). The material is based upon work supported by NASA under award number 80GSFC24M0006 (M.S.). R.I. was partially supported for this work by NSF AAG award 2009624. T.W. acknowledges financial support from the University of Illinois Vermilion River Fund for Astronomical Research. R.I.Y. was a JSPS grant fellow DC1 (2022--2025). This work was also supported by NAOJ ALMA Scientific Research Grant Code 2023-25A(P.I. H. Sano).

\end{acknowledgments}

\bibliographystyle{aasjournal}
\bibliography{IOPEXPORT_BIB}{}

\appendix
\renewcommand{\thefigure}{A\arabic{figure}}
\setcounter{figure}{0} 
\section{Large scale view of the LMC}
\begin{figure}[htb!]
    \centering
    \includegraphics[width=0.8\columnwidth]{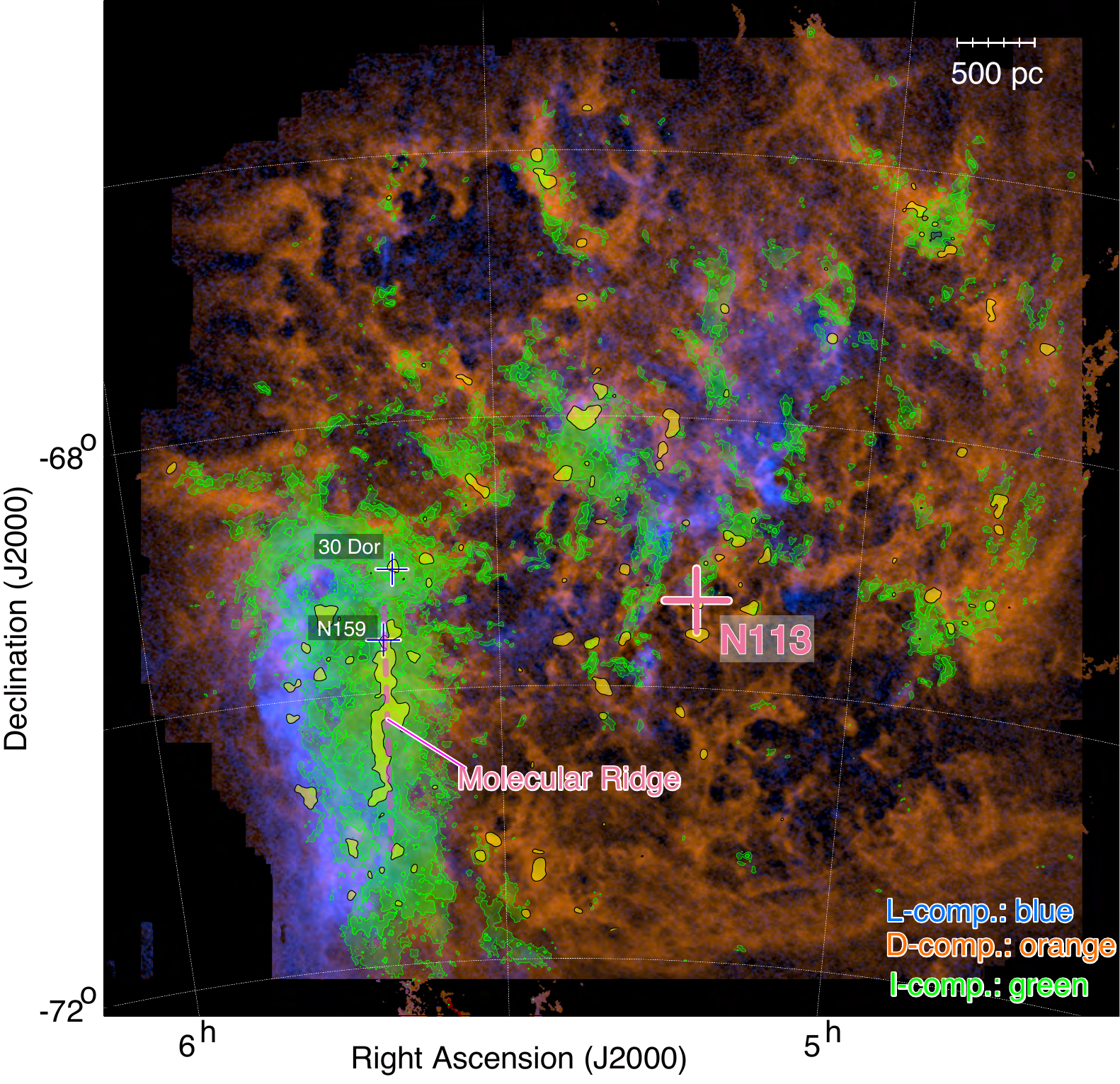}
    \caption{Three-color composite image of the whole LMC. Blue, orange, and green represent the L-component, D-component, and I-component, respectively.}
    \label{fig:LMC_whole}
\end{figure}

We focus on the massive star-forming region N113 in the LMC. Using high spatial resolution data obtained with ALMA+APEX, we derived the physical properties of CO clouds. To interpret these properties, we also analyzed the large-scale H$\;${\sc i} kinematics. As comparison targets, we selected the 30 Dor and N159 regions located in the molecular ridge of the LMC, where CO clouds have been studied in detail. The positions of these objects are plotted in Figure \ref{fig:LMC_whole}.

\section{Comparison of the MOPRA $\twelvecol$ data and APEX $\twelvecoh$ data}\label{sec:appA}
\begin{figure}[htb!]
    \centering
    \includegraphics[width=1.0\columnwidth]{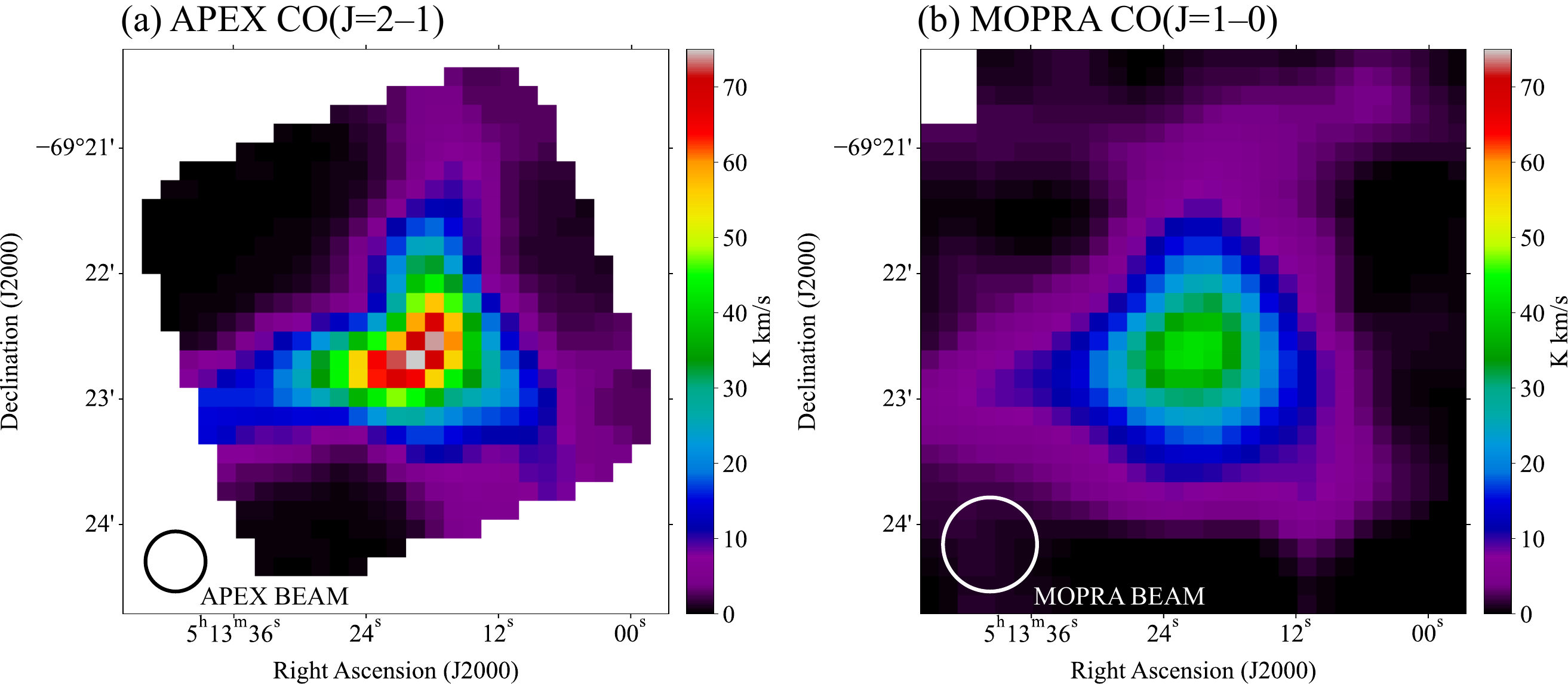}
    \caption{(a) Integrated intensity map of the $\twelvecoh$ emission obtained with APEX. (b) Integrated intensity map of the $\twelvecol$ emission obtained with MOPRA. The integration velocity range is (-10, 10)~km~s$^{-1}$ in the $V_\mathrm{offset}$ frame for both panels a and b.}
    \label{fig:sdcomp}
\end{figure}

\begin{figure}[htb!]
    \centering
    \includegraphics[width=0.5\columnwidth]{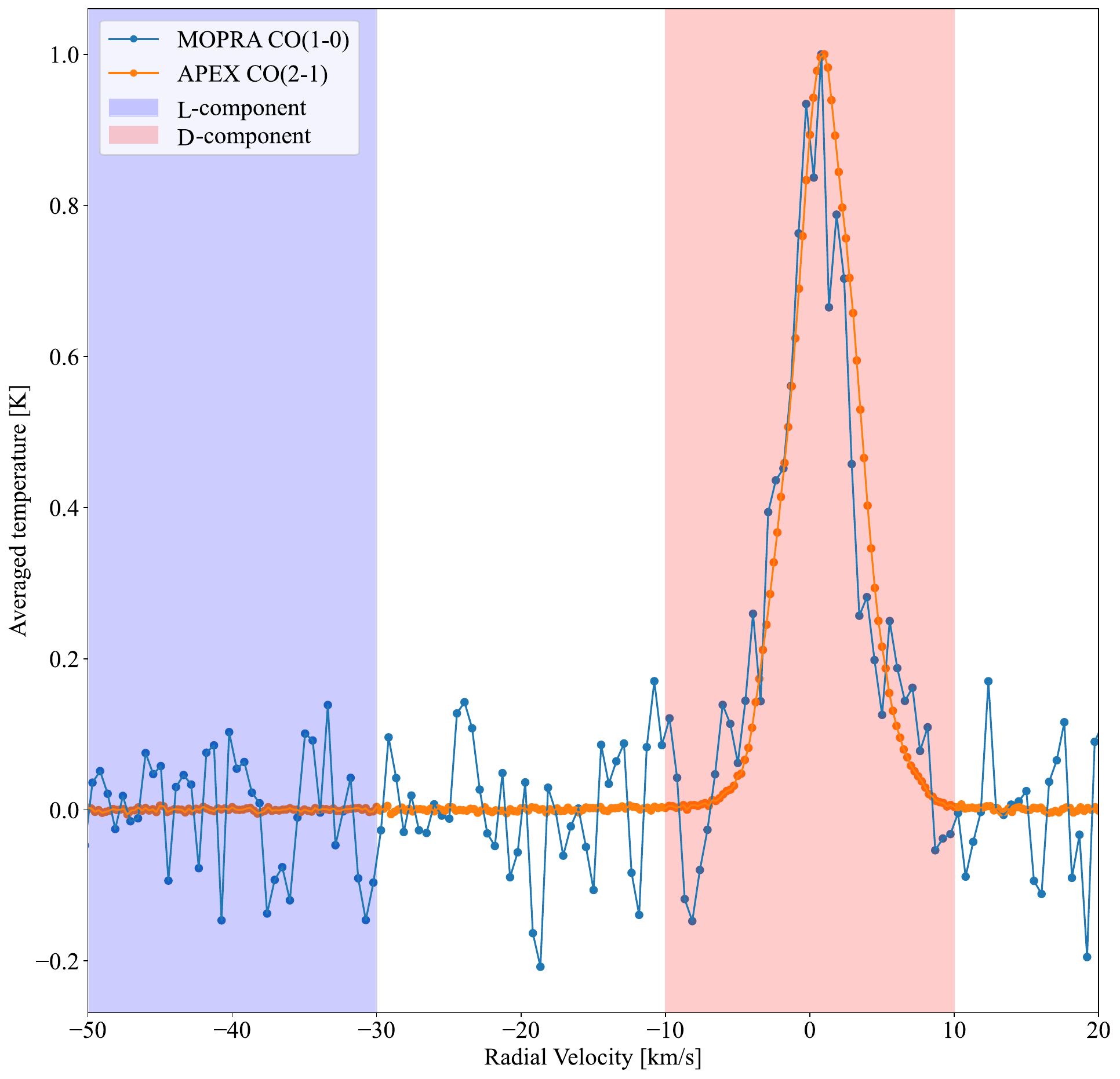}
    \caption{Normalized spectrum of N113 GMC in $\twelvecoh$ and $\twelvecol$ obtained with APEX and MOPRA, respectively. For the APEX data, we derived the average spectrum over the entire observed field.
For the MOPRA data, only the area overlapping with the APEX coverage was extracted, and the average spectrum within this region was obtained.
Both spectra were normalized such that their peak intensities are unity. The red and blue transparent belts indicate the velocity range of the D-component and L-component, respectively. The two CO emission lines are located in the velocity range of the D-component.}
    \label{fig:spcomp_sd}
\end{figure}

In this study, the large-scale CO distribution is traced using the MOPRA $\twelvecol$ data having a resolution of 45$''$, while the detailed structure toward N113 is investigated with the ALMA+APEX $\twelvecoh$ data with a resolution of $\sim$1$''$. To verify that there is no significant difference in their spatial or velocity distributions, we compared the APEX $\twelvecoh$ data ($\sim$28$''$ resolution) before combining with ALMA to the MOPRA $\twelvecol$ data, as shown in Figures \ref{fig:sdcomp} and \ref{fig:spcomp_sd}.

As seen in Figure \ref{fig:sdcomp}, the spatial distributions appear slightly different due to the difference in spatial resolution, but the peak positions and overall cloud extents are consistent considering the beam sizes. The velocity distributions also show the same trend, indicating that both datasets trace the same velocity component associated with the D-component. The CO cloud is located in the velocity range of the D-component, indicating that major precursor of the cloud is D-component.

\end{document}